\documentclass[aps,prb,twocolumn,superscriptaddress,bibliography]{revtex4-2}

\usepackage{amsfonts}
\usepackage{mathrsfs}
\usepackage{amsmath}
\usepackage{color}
\usepackage{graphicx}
\usepackage{bm}
\usepackage{amssymb}
\usepackage{xspace}
\usepackage{epstopdf}
\usepackage{dcolumn}
\usepackage{longtable}
\usepackage{multirow}
\usepackage{float}
\usepackage{comment}
\usepackage[bottom]{footmisc}

\usepackage[colorlinks=true, letterpaper=true, pdfstartview=FitV,  linkcolor=blue, citecolor=blue, urlcolor=blue]{hyperref}

\begin{document}

\title{Triple-well ferroelectricity and kagome-like Chern flat band in two-dimensional multiferroic CuVP$_2$Se$_6$}
\author{Brian Anchico}
\author{Jingyi Duan}
\email{duanjy0518@gmail.com}
\author{Haojie Sun}
\author{Minjun Wang}
\affiliation{ Centre for Quantum Physics, Key Laboratory of Advanced Optoelectronic Quantum Architecture and Measurement (MOE), School of Physics, Beijing Institute of Technology, Beijing, 100081, China }
\author{Mikhail Talanov}
\affiliation{ Center for Photonics and 2D Materials, Moscow Institute of Physics and Technology, Dolgoprudny 141700, Russia }
\author{Wei Jiang}
\email{wjiang@bit.edu.cn}
\affiliation{ Centre for Quantum Physics, Key Laboratory of Advanced Optoelectronic Quantum Architecture and Measurement (MOE), School of Physics, Beijing Institute of Technology, Beijing, 100081, China }
\affiliation{International Center for Quantum Materials, Beijing Institute of Technology, Zhuhai, 519000, China}

\date{\today}

\begin{abstract}
Two-dimensional multiferroics that host nontrivial topological bands offer a rich platform for correlated and tunable quantum phenomena, yet such materials remain rare. Here, using first-principles calculations, we reveal that monolayer CuVP$_2$Se$_6$ unites a tunable triple-well ferroelectric transition with a spin-polarized Chern flat band. The ferroelectric and paraelectric phases are close in energy and can be reversibly switched by moderate strain or an electric field. During the transition, a kagome-like flat band emerges near the Fermi level, which we describe via a minimal three-orbital tight-binding model on a triangular lattice. Furthermore, the system exhibits sizable magnetic anisotropy and a magnetization-dependent Chern insulating state: the Chern number is $C = \pm 1$ for out-of-plane magnetization but becomes trivial when the moments rotate in-plane. These findings establish CuVP$_2$Se$_6$ as a promising candidate for exploring electrically tunable flat-band correlations and topological magnetism in a multiferroic monolayer.
\begin{description}
\item[Keywords]
Two-dimensional, Multiferroics, Chern Flat band, Topological states, First-principles calculations

\end{description}
\end{abstract}


\maketitle

\section{\label{sec:level1}INTRODUCTION}
In the past decade, two-dimensional ferroelectrics have emerged as a vital class of functional materials. Numerous systems have been discovered, exhibiting polarization from diverse mechanisms across a broad range of Curie temperatures and spontaneous polarizations~\cite{Review2D_Discov_2_2022, Review2D_Discov_1_2023}. Their unique properties enable widespread applications in photovoltaics, sensing, actuation, tunnel junctions, transistors, and high-density nonvolatile memories~\cite{Review2D_Appl_1_2024, Photovoltaic2D_MoS2_2022, Functional2D_DomainWall_2017, In2Se3_Piezoelectric_2018, HybridPerovskite_UC_IPFE_Strain_2018}. Beyond standalone ferroelectricity, the coupling with magnetism in multiferroic materials introduces multiple symmetry breakings and rich magnetoelectric interactions~\cite{NiI2_SingLay_Type_II_Multif_2022}. A prominent example is the recently proposed type-III multiferroic TiCdO$_4$, which promises an exceptionally large magnetoelectric response~\cite{TiCdO4_Type-III_MonoLay_2025}. Furthermore, phase transitions in 2D materials are highly dynamic, often revealing novel fundamental phenomena that warrant deeper exploration~\cite{Review2D_PhaseTran_2021}.

A paradigmatic system in this field is the van der Waals (vdW) layered ferroelectric CuInP$_2$S$_6$, which has been synthesized down to the monolayer limit~\cite{CuInP2S6_Exp_vdW_2015, CuInP2S6_Exp_Ultrathin_2016, CuInP2S6_Exp_MultipleCuSites_2024,Synthetized_Chalcogenides_2023}. It exhibits size- and layer-dependent polarization~\cite{CuInP2S6_Theo_vdW_Layer_Order_2018, CuInP2S6_Exp_vdW_Polar_Order_2022}, tunable ground states under strain~\cite{CuInP2S6_Theo_MonoLay_Strain_2019}, and complex energy landscapes with more than two potential wells as a function of strain and temperature~\cite{CuInP2S6_TheoExp_vdW_QuadrupleWell_2020, CuInP2S6_Theo_vdW_StrainTemp_Wells_2023}. A significant extension of this family involves replacing the non-magnetic indium ion with a magnetic transition metal such as vanadium or chromium, yielding materials like CuVP$_2$Se$_6$ and CuCrP$_2$S$_6$. This substitution introduces magnetism, transforming them into multiferroics that retain the rich ferroelectric physics of their parent compound while enabling additional magnetoelectric couplings~\cite{TMPC_Theo_MultipleLayers_2022,TMPC_ExpTheo_SpinInduced_Ferroelectric_2022, TMPC_Theo_RashbaElectrically_2020}. While these materials have also been synthesized experimentally~\cite{CuCrP2S6_Exp_Synt_4UC_2023, CuCrP2S6_Exp_Synt_RT_4UC_2023, CuVP2S6_Exp_vdW_2025, CuVP2S6_CuCrP2S6_Exp_Synt_Flakes_2024}, a comprehensive understanding of their monolayer properties remains limited, motivating the present investigation~\cite{TMPC_Theo_Multiferroic_2018, TMPC_Theo_ValleySplitting_Topo_2022}.

Concurrently, the study of crystalline flat bands has seen remarkable theoretical and experimental progress~\cite{Jiang2019alieb,duan2024cataloging,Duan2024three,jiang2021giant,Jiang2021exotic,regnault2022catalogue,neves2024crystal,ye2019haas,kang2020dirac,liu2020orbital,wang2018large,li2021dirac,wu2025flat}, such as in experimentally synthesized kagome-lattice systems Fe$_3$Sn$_2$~\cite{ye2019haas}, FeSn~\cite{kang2020dirac}, CoSn~\cite{liu2020orbital}, Co$_3$Sn$_2$S$_2$~\cite{wang2018large}, YMn$_6$Sn$_6$~\cite{li2021dirac}, and CsCr$_3$Sb$_5$~\cite{wu2025flat}. These materials frequently host Chern flat bands, which are of great interest for realizing exotic correlated and topological phases\cite{liu2013flat,li2018realization,jiang2019dichotomy,Jiang2021exotic}.
However, multiferroic materials that combine ferroelectricity, magnetism, and spin-polarized Chern flat bands remain exceedingly rare~\cite{wang2025breathing,regmi2022spectroscopic}. The exploration of such multifunctional flat-band systems in experimentally viable materials thus presents a compelling research frontier.

In this work, we investigate multiferroic phase transition and emergent flat-band physics in monolayer CuVP$_2$Se$_6$. We first demonstrate a paraelectric-ferroelectric (PE-FE) phase transition governed by a triple-well potential, where the FE phase is metastable but can be stabilized by an external electric field or strain. Using nudged elastic band (NEB) calculations and first-principles methods, we fit a Landau free-energy expansion for the order parameter. Remarkably, we find that a kagome-like, spin-polarized flat band emerges that is robust against the PE-FE transition. We construct an effective tight-binding model based on the $d_{xy}, d_{x^2-y^2}, d_z^2$ orbitals on a triangular lattice to capture this physics and characterize the topological properties as a function of the magnetization orientation. Our work suggests that such multiferroic flat-band systems could host novel correlation-driven phenomena and enable innovative functional applications.

\section{\label{sec:level3} Calculation Methods}

The calculations were carried out within density functional theory using the Perdew–Burke–Ernzerhof (PBE) exchange–correlation functional~\cite{PBE_GGA_1996} and the projector-augmented-wave (PAW) method~\cite{PAW_method_1999}, as implemented in the \textsc{VASP} 5.4.4. A plane-wave energy cutoff of 500 eV was used, and the Brillouin zone was sampled with an $8\times8\times1$ $\Gamma$-centered Monkhorst–Pack \textit{k}-point mesh. A vacuum region exceeding 20~\AA\ was included along the out-of-plane direction to avoid spurious interections between periodic images. The on-site Coulomb interaction for the localized $d$ electrons were treated using the Dudarev approach~\cite{U-Value_Dudarev_1998}. Several van der Waals (vdW) correction schemes were tested~\cite{IVDW10_Grimme_2006, IVDW11_Grimme_2010, IVDW12_Grimme_2011}, including those used in previous studies of related materials~\cite{TMPC_Theo_Multiferroic_2018, CuInP2S6_Theo_MonoLay_Strain_2019}. However, the calculated potential barriers and ground states showed noticeable dependence on the choice of vdW method (see Appendix~\ref{vdww}), likely due to differences in pairwise interactions and damping functions optimized for other material classes~\cite{klimevs_vdW_Review_2012}.Given that vdW interactions decay as $\sim 1/r^6$ and the large vacuum spacing renders their effect negligible for the monolayer properties studied here, explicit vdW corrections were omitted in the main calculations. The Grimme approach with zero-damping~\cite{IVDW11_Grimme_2010} was confirmed to have a minimal impact on the potential barriers (Appendix~\ref{vdww}). Dipole corrections were included in all calculations~\cite{Dipole_Correction_1992}, while negligible in the absence of an external field, they become essential when an electric field is applied.

The nudged elastic band (NEB) method was employed using the \textsc{VASP} implementation, which incorporates the foundational ideas of Mills \textit{et al.} within transition-state theory~\cite{NEB_Mills_Jonsson_1995}. Intermediate images along the reaction pathway were generated using the VTST tools. The magnetic anisotropy energy (MAE) was calculated using a 40-atom supercell by imposing ferromagnetic spin orientations along different directions in the $XY$ and $XZ$ planes, with angular intervals of $30^\circ$ between successive configurations. For these MAE calculations, the on-site Hubbard $U$ correction was not included.

For the analysis of electronic properties, symmetry representations were obtained using \textsc{MSGCorep}~\cite{liu2023msgcorep} and \textsc{SpaceGroupIrep}~\cite{liu2021spacegroupirep}. A symmetry-adapted tight-binding model was constructed with~\textsc{MagneticTB}~\cite{zhang2022magnetictb}. Maximally localized Wannier functions were generated via \textsc{Wannier90} package~\cite{mostofi2008wannier90, mostofi2014updated}, and topological properties were calculated using the \textsc{WannierTools} package~\cite{wu2018wanniertools}. For electronic structure calculations requiring high \textit{k}-space resolution, a $\Gamma$-centered mesh with spacing finer than 0.01 Å$^{-1}$ was employed.

\section{\label{sec:level4} Crystal structures}

\begin{figure}[ht]
\includegraphics[width=\linewidth]{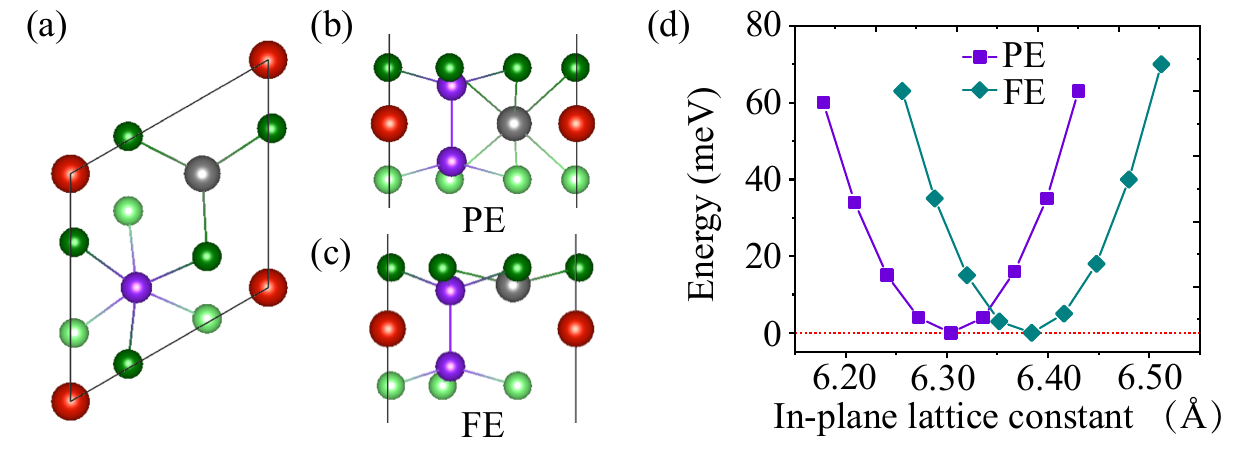}
\caption{Crystal structure of CuVP$_2$Se$_6$. (a) Top view. (b,c) Side views of the paraelectric and ferroelectric phases, respectively. (d) Schematic illustration of the energy variation under applied strain. Cu, V, and P atoms are represented by gray , red, and purple spheres, respectively, while dark-green and light-green spheres denote the upper and lower Se atoms.}
\label{fig:CuVP2Se6_Structure_Strain}
\end{figure}

The monolayer structure of $\mathrm{CuVP_2Se_6}$ [Fig.~\ref{fig:CuVP2Se6_Structure_Strain}(a)] can be described as a network of nearly octahedral Se cages, with Cu and V ions occupying the octahedral sites. These cations are coordinated by the $(\mathrm{P_2Se_6})^{4-}$ units, which consist of a P–P dimer sandwiched between two triangular $\mathrm{Se_3}$ layers. The paraelectric (PE) to ferroelectric (FE) phase transition does not significantly alter the in-plane geometry of these triangular layers; instead, it involves a collective rotation or tilting of the polyhedra, resembling the mechanism of a coloring-triangle lattice~\cite{Coloring-triangle_lattice_2019}.

In the high-temperature PE phase ($T>T_c$), which belongs to the space group P312 (SG~149), the Cu electron density is symmetrically distributed between top and bottom sites, analogous to a centered Cu position on average [Fig.\ref{fig:CuVP2Se6_Structure_Strain}(b)]. This configuration, also observed experimentally in layered analogs~\cite{CuVP2S6_Exp_Layer_1990, CuInP2S6_Exp_Ferrielectric_1997, CuCrP2S6_Exp_Synt_RT_4UC_2023}, results in zero net spontaneous polarization. Upon cooling below $T_c$, the system transitions to the FE phase with space group P3 (SG~143) [Fig.~\ref{fig:CuVP2Se6_Structure_Strain}(c)]. Here, the Cu electron density localizes preferentially at either the top or bottom site, breaking inversion symmetry and generating a finite spontaneous polarization.

To assess the phase stability, we applied isotropic in-plane strains from $-2\%$ (compressive) to $+2\%$ (tensile) relative to the relaxed structure and performed full structural relaxation. We find a remarkably small energy difference between the PE and FE phases, indicating their possible coexistence under ambient conditions. The phase stability is highly strain-tunable: tensile strain systematically favors the FE phase, while compressive strain stabilizes the PE phase  [Fig.~\ref{fig:CuVP2Se6_Structure_Strain}(d)]. This sensitive response contrasts with earlier studies on related compounds (e.g., CuInP$_2$S$_6$), where a larger PE–FE energy barrier required strains exceeding $\pm2\%$ to switch the ground state~\cite{CuInP2S6_Theo_MonoLay_Strain_2019, TMPC_Theo_Multiferroic_2018,CuInP2S6_Exp_Stress_2023}. The small barrier predicted here suggests that monolayer CuVP$_2$Se$_6$ is intrinsically more tunable by strain, as well as by temperature, electric fields, and other external stimuli discussed below. We note that the ground state of this material is antiferroelectric (Appendix~\ref{App:AFE_GS}), in line with most TMPCs; nevertheless, a metastable FE phase has been experimentally stabilized via tip voltage and electron-beam irradiation~\cite{CuVP2S6_Exp_vdW_2025, CuCrP2S6_Exp_Synt_RT_4UC_2023}.

\section{\label{sec:level5}PARAELECTRIC TO FERROELECTRIC TRANSITION}

\begin{figure}[ht]
\includegraphics[width=\linewidth]{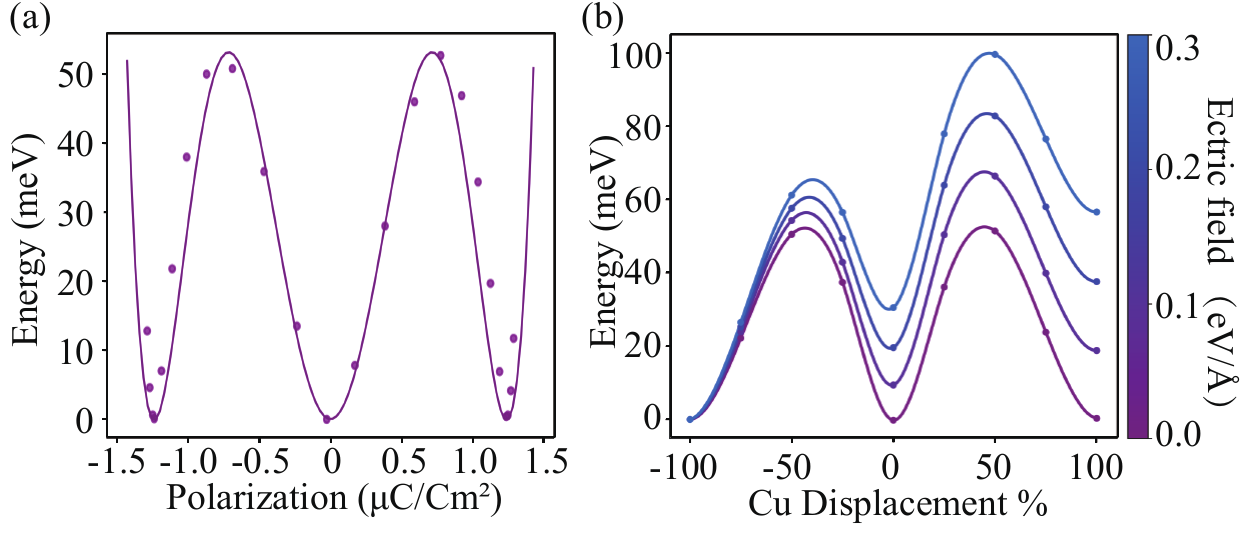}
\caption{Energy-polarization diagram and Landau free energy fitting. (a) Energy–polarization diagram illustrating the complete polarization switching process. The points represent data obtained from the NEB calculations, while the solid curve is obtained by fitting the data to Eq.~(\ref{eq:landau_exp}), with fitting parameters $\alpha$ = 467 meV$(\mu C/cm^2)^{-2}$, $\beta$ = -1217 meV$(\mu C /cm^2)^{-4}$, and $\gamma$ = 594 meV$(\mu C /cm^2)^{-6}$. (b) Evolution of the potential barriers between the paraelectric and ferroelectric phases under an applied electric field calculated from NEB with E-field (points) and fittings (lines).}
\label{fig:NEB}
\end{figure}

To characterize the PE-FE phase transition and map the switching pathway between the two polarized states, ($\mathrm{-P_0}$ and $\mathrm{+P_0}$), we performed nudged elastic band (NEB) calculations. The initial image corresponds to the Cu atom located near the top $\mathrm{Se_3}$ layer [Fig.~\ref{fig:CuVP2Se6_Structure_Strain}(c)], the intermediate images pass through the PE configuration [Fig.~\ref{fig:CuVP2Se6_Structure_Strain}(b)], and the final image places Cu atom near the bottom $\mathrm{Se_3}$ layer. Along the pathway, the in-plane lattice constant expands gradually from the PE to the FE phase [Fig.~\ref{fig:CuVP2Se6_Structure_Strain}(d)]; in the NEB calculation, the in-plane lattice constant of each image was allowed to relax accordingly. The resulting energy profile is plotted in Fig.~\ref{fig:NEB}(a).

The landscape exhibits a triple-well potential, a fingerprint of a first-order phase transition. The system resides in close proximity to its phase coexistence region, as confirmed by the minimal energy difference between the FE and PE states (Table \ref{tab:U_Value}). This energy quasi-degeneracy implies that phase transitions can be driven by minimal external stimuli. Specifically, a small electric field can induce the PE-to-FE transition and subsequently switch the polarization within the FE state, which should result in a pronounced dielectric susceptibility and a low coercive field. The overall shape of the energy profile is consistent with earlier reports~\cite{TMPC_Theo_Multiferroic_2018}, though the barrier heights differ. Similar energy profiles are obtained when including vdW corrections (Appendix~\ref{vdww}) or an on-site Hubbard $U$ parameter (Table \ref{tab:U_Value}).

\begin{table}[h] 
  \caption{Energy of each phase under electron--electron interaction (U value), in units of eV. The numbers in each column denote the expansion of the unit cell along the three crystallographic directions ($a$, $b$, and $c$); thus, 111 corresponds to the 10-atom unit cell, 121 to a 20-atom supercell. } 
  \label{tab:mi_tabla} 
  \centering 
  \begin{tabular}{cccccc} 
    \hline 
      & FE & PE & FE & PE & FE  \\ 
     U & FM 111 & FM 111 & FM 121 & FM 121 & AFM 121  \\     
\hline 
0 & \textbf{-46.851} & \textbf{-46.851} & -93.762 & -93.729 &  -93.748\\ 
1 & -46.045 & {-46.093} & -92.283 & -92.223 &   -92.276  \\
2 & -45.308 & {-45.406} & -90.924 & -90.833 & -90.916 \\ 
3 & -44.644 & {-44.768} & {-89.663} & -89.537  & -89.660 \\ 
4 & -44.080 & {-44.413} & -88.492 & {-88.827} &  -88.489  \\ 
    \hline 
  \end{tabular}
  \label{tab:U_Value}
\end{table}

Within the Landau theory of phase transitions, the displacive transition between the PE phase with space group P312 and the FE phase with space group P3 is described by an order parameter transforming according to the one-dimensional irreducible representation $\Gamma$2 of the parent phase's space group P312. The transformation properties of this critical order parameter are identical to those of the polarization vector along the z-axis, which confirms the proper ferroelectric nature of the transition. Based on this symmetry analysis and the first-order character evidenced by the triple-well profile, the simplest form of the thermodynamic potential is:
\begin{equation}
F(P,T)=\frac{1}{2}\alpha P^2+\frac{1}{4}\beta P^4+\frac{1}{6}\gamma P^6 -EP
\label{eq:landau_exp}
\end{equation}
The fitted coefficients in Equation (\ref{eq:landau_exp}) have the following signs: $\alpha$ > 0, $\beta$ < 0, and $\gamma$ > 0. The positive sign of $\alpha$ is standard, the negative sign of $\beta$ is the hallmark of a first-order transition, and the positive sign of $\gamma$ is required for stability.

We further explore how an external electric field $E$ modifies the energy landscape via the linear $EP$ coupling in Eq.~(\ref{eq:landau_exp}). As $|E|$ increases, the PE well becomes shallower, and one FE minimum deepens into the global ground state [Fig.~\ref{fig:NEB}(b)]. Beyond a critical field $E_c$, the PE well vanishes, leaving a single FE minimum. This is also verified from our first-principles calculations [dots in Fig.~\ref{fig:NEB}(b)]. It is important to note that, for this material, the PE phase is thermodynamically stable for $T>T_c$. Upon cooling, the system may remain trapped in the PE phase unless external excitations or engineered perturbations (e.g., by strain and/or electric field) induce the transition to the FE phase. 

Using both a classical point-charge model and the Berry-phase formalism~\cite{Vanderbilt_King_ModernPolarization_1993, Resta_ModernPolarization_1994, Spaldin_ModernPolarization_2012}, we estimate the out-of-plane spontaneous polarization at zero field to be $1.24~\mu\mathrm{C/cm}^2$ and $1.07~\mu\mathrm{C/cm}^2$, respectively. These values are about half of the experimental polarization reported for five-layer CuInP$_2$S$_6$~\cite{Review2D_Discov_1_2023}, and consistent with the known layer-dependent reduction in polarization—e.g., in CuInP$_2$Se$_6$, the polarization drops by roughly a factor of two from four layers to a monolayer~\cite{CuInP2Se6_Polarization_BulkMono_2017}.

\section{magnetic and electronic properties}

\subsection{Magnetic properties}

\begin{figure}[htbp]
    \centering
    \includegraphics[width=1\linewidth]{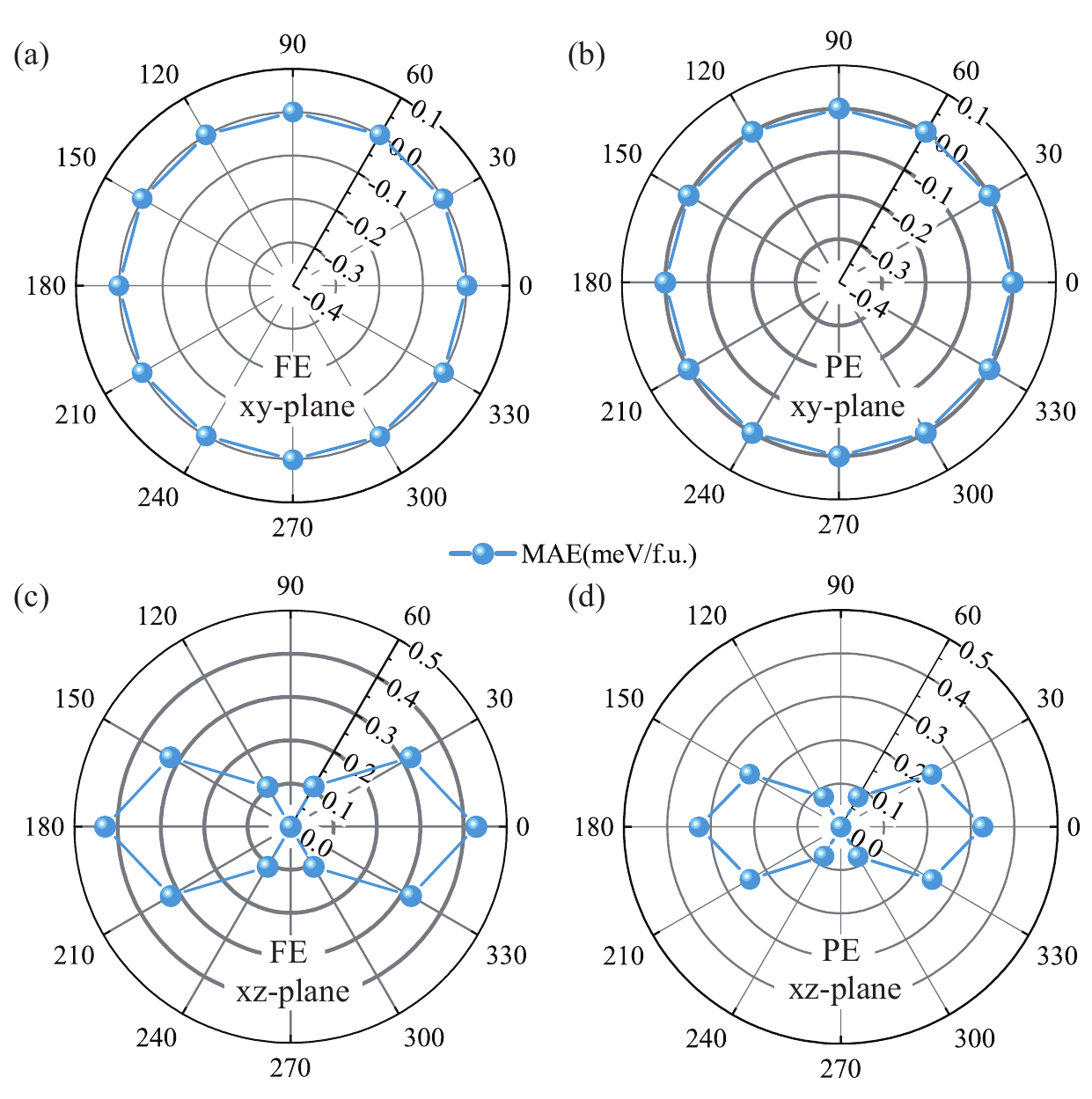}
    \caption{Contour maps of the magnetic anisotropy energy (MAE). (a) and (b) MAE for the xy-plane for the ferroelectric and paraelectric (PE) phases, respectively. (c) and (d) Same as (a,b) for the xz-plane. The energy zero is set to the lowest energy of the system in each situation. MAE is given in meV per formula unit.}
    \label{fig:MSG}
\end{figure}

We next investigate the magnetic properties of monolayer CuVP$_2$Se$_6$. To determine the magnetic ground state, we compare the energies of ferromagnetic (FM) and antiferromagnetic (AFM) orderings for both FE and PE phases in a 1$\times$2$\times$1 supercell. This comparison is performed for several values of the on-site Hubbard $U$ parameter, with the results summarized in Table~\ref{tab:U_Value}). For the FE phase, the FM configuration is the most stable across all tested $U$ values, indicating a robust magnetoelectric coupling that favors ferromagnetism in the polarized state. In the PE phase, while FM order is easily stabilized, AFM states are hard to converge, suggesting their instability. The magnetic anisotropy energy (MAE) was also calculated, as presented in Fig.~\ref{fig:MSG}. Consistent with the crystal symmetry, all orientations within the xy-plane are energetically degenerate due to the in-plane threefold rotational (C$_3$) symmetry [Figs.~\ref{fig:MSG}(a)-(b)]. In the xz-plane, the energy varies smoothly as the magnetization rotates from the x-axis toward the z-axis, as shown in Figs.~\ref{fig:MSG}(c)-(d). For both FE and PE phases, the easy axis lies along the z-axis—either parallel or antiparallel to the polarization direction. The MAE is slightly larger in the PE phase than in the FE phase, reflecting subtle differences in their electronic environments.

\subsection{Electronic Flat band}

\begin{figure}[htb]
    \centering
    \includegraphics[width=1\linewidth]{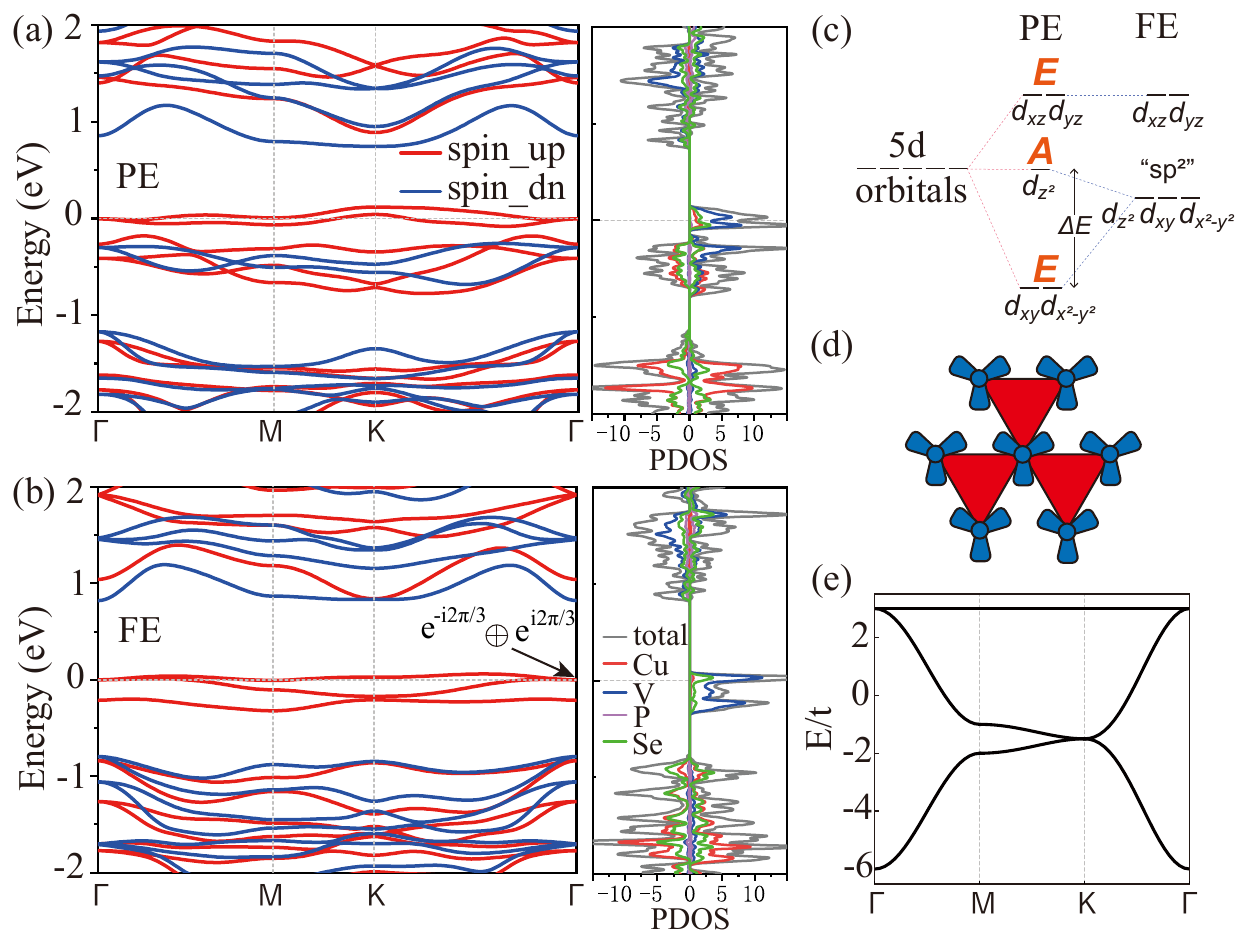}
    \caption{Electronic properties of CuVP$_2$Se$_6$.
Band structure and elemental projected density of states of (a) PE phase and (b) FE phase; (c) Splitting of $d$ orbitals in the octahedral crystal field, $C_3$ point group, and FE phase.
(d) a schematic diagram of the ${s, p_x, p_y}$ triangular model, and (e) the corresponding flat band structure when $\epsilon = 0$ and $t = 1$.}
    \label{P4}
\end{figure}

The switching between triple-well states from -FE to PE to +FE not only governs the macroscopic polarization but is also expected to significantly inpact the electronic structure. We therefore examine the electronic properties of FM CuVP$_2$Se$_6$ in both the PE and FE phases. Without spin-orbit coupling (SOC), both phases exhibit a semimetallic character, with one spin channel conductive (displaying band crossing at the Fermi level) and degenerate at the $\Gamma$ point, while the other remains insulating, as shown in Fig.~\ref{P4}(a,b). The transition from PE to FE markedly reshapes the band structure, most strikingly, it introduces kagome-like flat and Dirac bands near the Fermi level in the FE phase and enlarges the spin gap from ~1eV in PE phase to ~2eV in FE phase. To understand the origin of those states near the Fermi level, we analyzed the projected density of states (PDOS). This reveals that V-derived orbitals dominate around the Fermi energy in both phases, whereas Cu-derived bands lie at deeper energies (approximately −2eV).

Given that the main changes between PE and FE is the migration of Cu ions, which can significantly alters the orbital hybridization on V atoms, thus resulting in changes in the band structure. Such orbital reconstruction can be understood from the crystal\mbox{-}field environment [Fig.~\ref{P4}(c)]. In the PE phase, the $d$-orbital splitting follows C$_3$ symmetry: $(d_{xy}, d_{x^2-y^2})$ and $(d_{xz}, d_{yz})$ form two degenerate pairs, while $d_{z^2}$ remains isolated. In the FE phase, the energy separation between $(d_{xy}, d_{x^2-y^2})$ and $d_{z^2}$ collapses ($\Delta E \approx 0$), and near $E_F$ these orbitals reconstruct into an effective ${s, p_x, p_y}$-like hybrid basis. Orbital\mbox{-}resolved band structures detailing this reconstruction are provided in Appendix~\ref{Orbit}.

As a consequence of this modified hybridization, not only does the spin splitting increase, but the bands near $E_F$ also become significantly less dispersive. A kagome\mbox{-}like motif emerges, characterized by a flat band touching a dispersive band at $\Gamma$ and a Dirac-like state at K. This flat-band feature is robust and only weakly influenced by the Hubbard U parameter (see Appendix \ref{fbdos}, Fig. \ref{fig:PDOS_U-value}). To the best of our knowledge, an explicit tight-binding realization of a kagome flat band on a triangular lattice has not been previously report, although models based on (p$_x$, p$_y$,$s$) orbitals on triangle lattice has been developed to exhibit quantum spin Hall states~\cite{Wang2016}. To explain the origin of this flat band, we construct a minimal tight\mbox{-}binding model on a triangular lattice using the ($d_{xy}, d_{x^2-y^2}, d_z^2$) orbital basis:

\begin{equation}
\ensuremath{\mathcal{H}}  =  \mathrm{\epsilon}\sum_{i}c^\dagger_{i} c_{i}+
\mathrm{t}\sum_{<i,j>}c^\dagger_{i} c_{j} +H.c,
\end{equation}
where $\epsilon$ is the on-site energy and $t$ the nearest-neighbor hopping. In the ideal case ($\epsilon= 0 $ and $t = 1$), the model reproduces the key features around E$_F$ observed in the first-principles band structure [Fig.~\ref{P4}(e)]. Moreover, because the FE phase inherently breaks inversion symmetry, the degeneracy at the $K$ point is lifted relative to the canonical kagome bands. Further details of the Hamiltonians are provided in Appendix~\ref{ham}. We note that this $d$ orbital model is mathematically equivalent to the ( $p_x$, $p_y$,$s$,) basis, and the description can be generalized to other equivalent orbital representations. The effects of strain, Hubbard $U$, and Cu displacement on the band structure are discussed systematically in Appendix~\ref{fbdos}.

\subsection{Topological properties}

\begin{figure}[htbp]
    \centering
\includegraphics[width=1\linewidth]{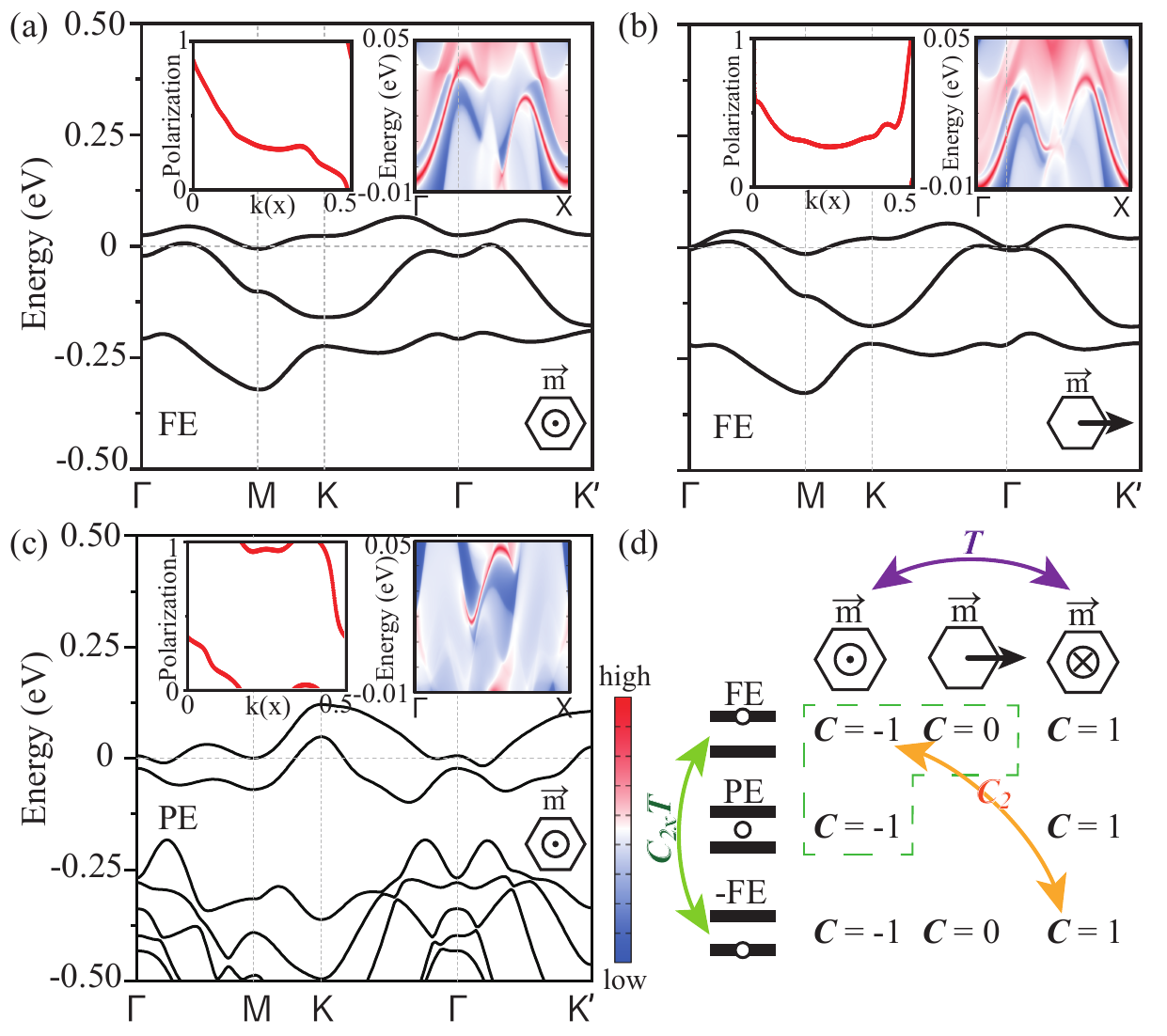}
    \caption{Topological properties of the CuVP$_2$Se$_6$  influenced by the magnetization direction and $\pm$FE-PE switching. (a) and (b) Band structure of FE phase considering spin-orbit coupling, with magnetic moment directions of $[001]$ and $[100]$. (c) Same as (a) for the PE phase with  $[001]$ magnetization direction. The insets in (a)-(c) represent the sum of the two occupied bands wannier charge centers (wcc) and edge states of the corresponding magnetic configurations. (d) Relationship between the symmetry operation of the PM, FE, and PE phases and the magnetization directions, and the corresponding Chern numbers for various configurations.}
    \label{topo}
\end{figure}

In this section, we investigate the topological properties of the system and their dependence on the magnetization direction. Without SOC, the bands are gapless at the $\Gamma$ point. Band-representation analysis shows that this degeneracy is formed by states with three-fold rotation eigenvalue $e^{\pm i2\pi/3}$ about the $z$ axis, bound together by the spin-group symmetry $[C_2|{\cal T}]$ (where $\cal T$ denotes time reversal), which acts as an effective time-reversal symmetry, as shown Fig.~\ref{P4}(b). When SOC is taken into account, the symmetry reduces from the spin group to a magnetic space group and $[C_2|{\cal T}]$ is broken. As a result, the degeneracy at $\Gamma$ is lifted and a finite band gap opens, allowing a well-defined topological character to emerge.

With SOC, the electronic and topological properties becomes sensitively to the magnetization direction. We first consider the case with magnetization along the $[001]$ direction. The system then belongs to the magnetic space group No. 143.1.1231, generated by $C_{3z}$, and its band structure is shown in Fig.~\ref{topo}(a). Consistent with symmetry, the degeneracy at $\Gamma$ is removed. The conduction and valence bands at $\Gamma$ carry $C_{3z}$ eigenvalues $e^{i\pi/3}$ and $e^{-i\pi/3}$, respectively. Once $[C_2|{\cal T}]$ is broken, the bands at the $K$ and $K'$ valleys are no longer degenerate, permitting a nonzero Chern number. Wilson-loop calculations confirm that the occupied bands have Chern number $C=-1$, as shown in the inset of Fig.~\ref{topo}(a).

We next examine the effect of in-plane magnetization by computing the band structure and Wilson loop for magnetization along $[100]$, as shown in Fig.~\ref{topo}(b). In this case, the Wilson-loop spectrum does not wind, indicating a topologically trivial state with Chern number $C=0$. Thus, as the magnetization rotates from $+m_z$ (out-of-plane) to $m_x$ (in-plane), the system undergoes a topological phase transition where the Chern number evolves from $-1$ to $0$. We also explore how the PE–FE transition affects topology. As shown in Fig.~\ref{topo}(c), when the magnetization is kept out-of-plane as in Fig.~\ref{topo}(a) (i.e., $+m_z$), the PE phase remains a Chern insulator with $C=-1$. This demonstrates that the Cu-atom displacement does not compromise the robustness of the topological phase.

The topological character of other symmetry-related configurations can be deduced without additional calculation, as summarized in Fig.~\ref{topo}(d). The $+m_z$ and $-m_z$ phases are related by time-reversal symmetry ${\cal T}$, which reverses the Chern numbers. The $\pm$FE phases are linked by the antiunitary symmetry $C_{2x}{\cal T}$, which preserves the Chern number. Combining these operations, all phases are connected by $C_2$ symmetry, allowing the Chern number of the remaining five configurations to be defined directly.

To further confirm the bulk-boundary correspondence, we computed edge spectra for three magnetization directions [Fig.~\ref{topo} and Appendix~\ref{tp}]. For $[001]$ and $[00\bar{1}]$, chiral edge states traverse the bulk gap in both $\pm$FE phases, propagating in opposite directions consistent with $C=-1$ and $C=+1$, respectively. In contrast, for in-plane magnetization along $[100]$, the edge spectrum is topologically trivial and no gap-crossing chiral modes appear. The PE phase similarly exhibits $C=\mp 1$ under $\pm m_z$ magnetization. These results demonstrate how the coupled switching of ferroelectric polarization and magnetization direction controls the topological state of the system.

\section{Conclusion}

In summary, we have identified monolayer CuVP$_2$Se$_6$ as a two-dimensional multiferroic exhibiting a characteristic triple-well potential. Its spontaneous polarization is comparable to that of the prototypical ferroelectric CuInP$_2$S$_6$, while the FE and PE states are highly tunable via moderate strain or external electric fields—a finding supported by our strain-dependent lattice analysis and NEB calculations of the switching pathway.  Crucially, the ferroelectric transition drives a reconstruction of the electronic structure, giving rise to kagome-like flat and Dirac bands near the Fermi level. Using a three‑orbital triangular‑lattice model, we explain the origin of these flat bands and demonstrate how the magnetization direction controls the topological state, yielding Chern insulators with $C=\pm 1$ for out‑plane magnetization and a trivial phase for in‑plane alignment.

Our results establish CuVP$_2$Se$_6$ as a promising platform for exploring correlated flat‑band physics and electrically tunable topology in a multiferroic setting. Several intriguing directions remain for future study. The interplay between electronic correlations and topological states warrants deeper investigation, particularly regarding possible correlated phases such as superconductivity, which has been linked to flat bands in other systems. The temperature dependence of the phase transition also calls for further computational and experimental characterization to fully map the phase diagram. Finally, the accessibility of metastable FE and topological states suggests opportunities for designing novel non‑volatile memory and logic devices that leverage both ferroelectric and topological switching.

\begin{acknowledgments}
This work is supported by the National Key R\&D Program of China (Nos. 2022YFA1403500, 2024YFA1408400), the National Natural Science Foundation of China (No. 12204037, No. 12547158) and the Beijing Institute of Technology Research Fund Program for Young Scholars. T.M.V. acknowledges support from the Russian Science Foundation Grant No. 25-42-00058.
\end{acknowledgments}

\appendix

\begin{table*}[h] 
\section{Van der Waals}\label{vdww}
  \caption{Total energies (in eV) of the ferroelectric (FE) and paraelectric (PE) phases of CuVP$_2$Se$_6$ and CuVP$_2$S$_6$ calculated using different van der Waals (vdW) correction schemes. All calculations correspond to ferromagnetic (FM) configurations. Although the inclusion of vdW corrections modifies the relative energy differences and, in some cases, the energetic ordering between the FE and PE phases, Grimme’s approach with a zero-damping function (IVDW=11)~\cite{IVDW11_Grimme_2010} has a negligible effect on the energy difference between the phases. The energetically favored ground state for each vdW scheme is highlighted in bold.} 
  \centering 
  \begin{tabular}{|c|c|c|c|c|} 
    \hline  
     vdW & CuVP$_2$Se$_6$-FE & CuVP$_2$Se$_6$-PE & CuVP$_2$S$_6$-FE & CuVP$_2$S$_6$-PE\\ 
    \hline 
    Non-IVDW & \textbf{-46.851} & \textbf{-46.851} & \textbf{-51.881} & -51.766 \\ 
    IVDW=10 & -48.588 & \textbf{-48.664} & \textbf{-53.305} & -53.274 \\ 
    IVDW=11 & \textbf{-48.531} & -48.528 & \textbf{-53.353} & -53.237 \\ 
    IVDW=12 & -49.512 &\textbf{ -49.662} & -54.393 & \textbf{-54.416} \\ 
    \hline 
  \end{tabular}

  \label{tab:mi_tabla}  
\end{table*}

\begin{figure*}[h]
\section{Antiferroelectric Ground state}\label{App:AFE_GS}
\centering
\includegraphics[width=0.6\linewidth]{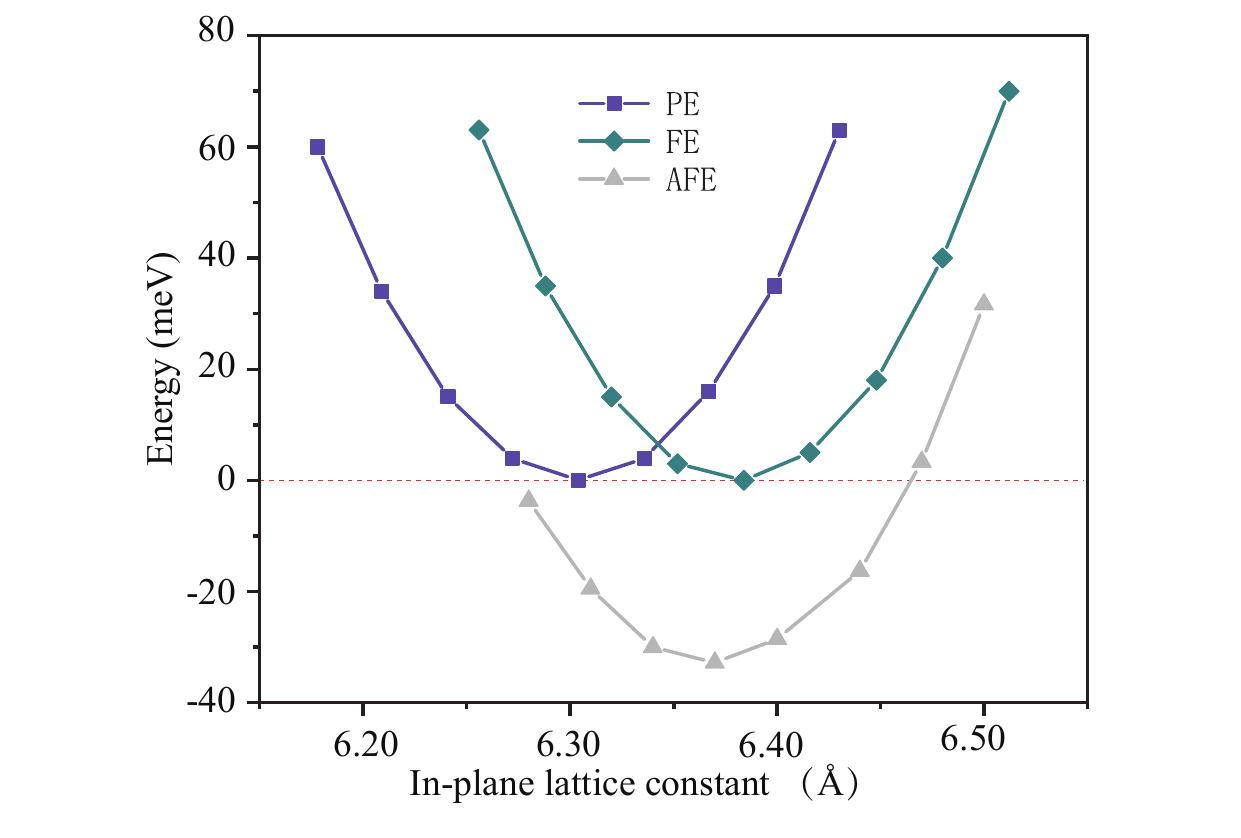} 
\caption{ Strain–energy diagram showing the antiferroelectric (AFE) ground state. The ferroelectric (FE) phase can be stabilized with a relatively small applied strain. The strain range considered is from $-2\%$ to $+2\%$.}
\label{Fig:AFE_GS}
\end{figure*}

\begin{figure*}[h]
\section{Orbital projected band structure} \label{Orbit}
    \centering
 \includegraphics[width=0.6\linewidth]{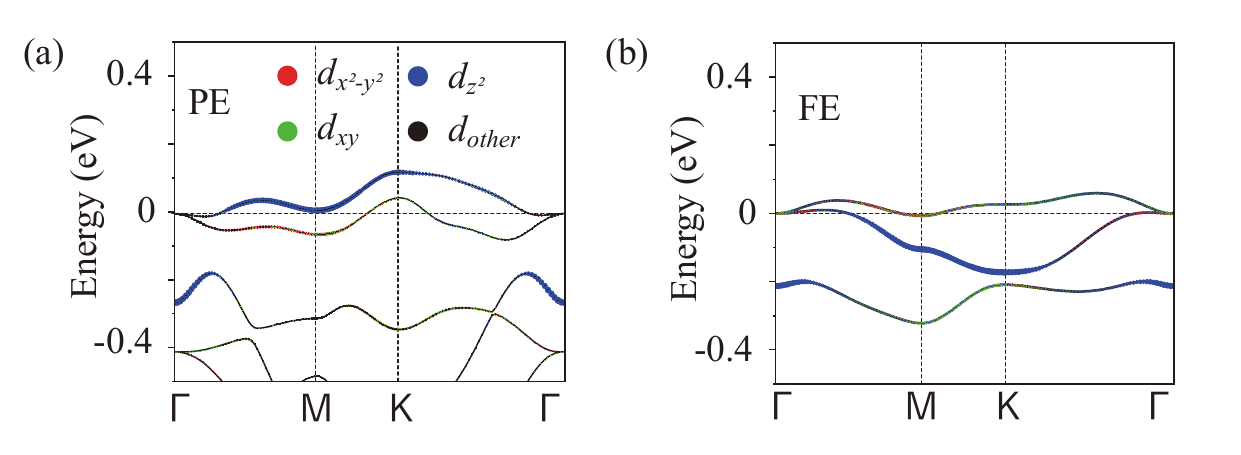}
    \caption{Orbital projected band structure of CuVP$_2$Se$_6$ (a) PE phase (b) FE phase.}
   
\end{figure*}

\begin{figure*}[h]
\section{Hamiltonian}\label{ham}

\begin{equation}
\ensuremath{\mathcal{H}} =  t
\begin{pmatrix}
H_{11} &   0  & H_{13} \\
  0 & H_{22} & H_{23} \\
 H_{13}^\dagger & H_{23}^\dagger & H_{33} 
\end{pmatrix},
\end{equation}

\begin{equation}
\begin{aligned}
H_{11} &= \tfrac{1}{2} e^{i(-k_x-k_y)}
       + \tfrac{1}{2} e^{i(k_x+k_y)}
       + \tfrac{1}{2} e^{-i k_x}
       + \tfrac{1}{2} e^{i k_x} \\
     &\quad + \tfrac{1}{2} e^{-i k_y}
       + \tfrac{1}{2} e^{i k_y} \\[6pt]
H_{13} &= \tfrac{\sqrt{3}}{2} e^{i(-k_x-k_y)}
       + \tfrac{\sqrt{3}}{2} e^{-i k_x} \\
     &\quad - \tfrac{\sqrt{3}}{2} e^{i k_x}
       - \tfrac{\sqrt{3}}{2} e^{-i k_y} \\[6pt]
H_{22} &= \tfrac{1}{2} e^{i(-k_x-k_y)}
       + \tfrac{1}{2} e^{i(k_x+k_y)}
       + \tfrac{1}{2} e^{-i k_x}
       + \tfrac{1}{2} e^{i k_x} \\
     &\quad + \tfrac{1}{2} e^{-i k_y}
       + \tfrac{1}{2} e^{i k_y} \\[6pt]
H_{23} &= \tfrac{1}{2} e^{i(-k_x-k_y)}
       - e^{i(k_x+k_y)}
       + \tfrac{1}{2} e^{-i k_x}
       + \tfrac{1}{2} e^{i k_x} \\
     &\quad + \tfrac{1}{2} e^{-i k_y}
       - e^{i k_y} \\[6pt]
H_{33} &= - e^{i(-k_x-k_y)}
       - e^{i(k_x+k_y)}
       - e^{-i k_x}
       - e^{i k_x} \\
     &\quad - e^{-i k_y}
       - e^{i k_y}
\end{aligned}
\end{equation}

\end{figure*}

\begin{figure*}[h] 
    \section{Band structure and DOS} \label{fbdos}
    \centering
    \begin{minipage}[b]{0.48\textwidth}
        \centering
        \includegraphics[width=\textwidth]{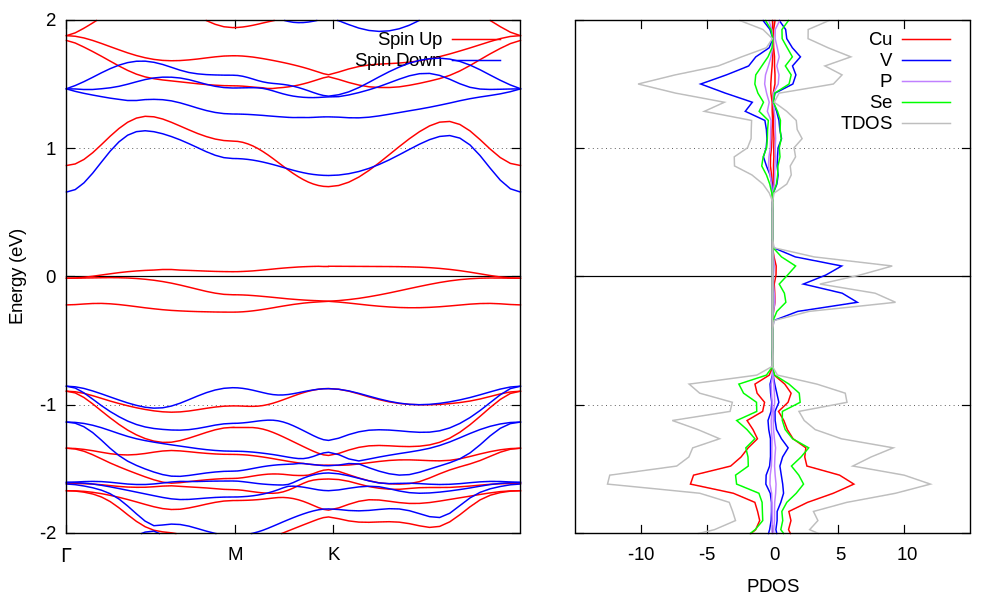} 
    \end{minipage}
    \hfill 
    \begin{minipage}[b]{0.48\textwidth}
        \centering
        \includegraphics[width=\textwidth]{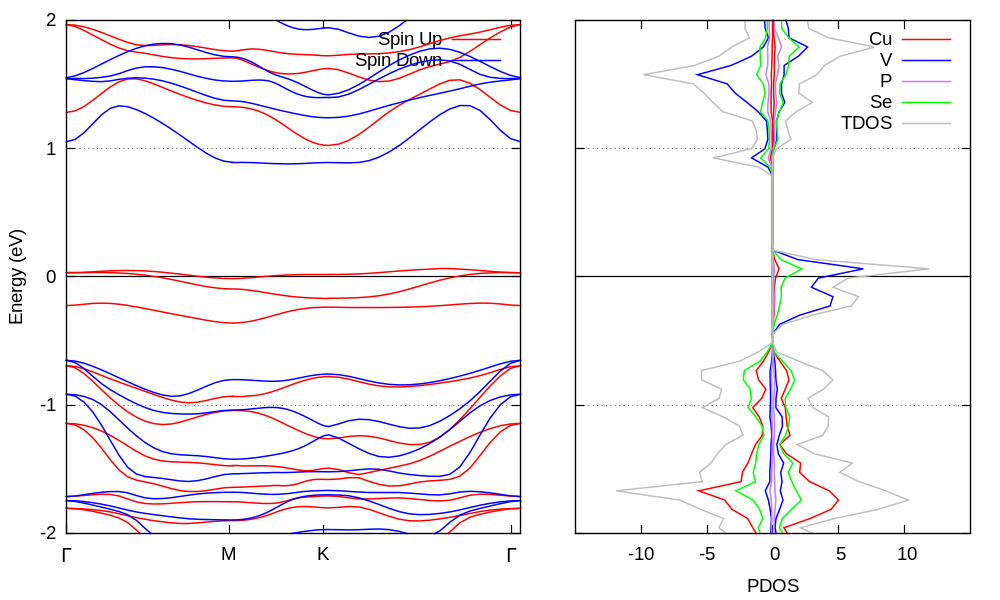}  
    \end{minipage}
    \caption{Band structure and projected density of states (PDOS) of the ferroelectric (FE) phase under 2\% compressive strain (left panel) and $-2\%$ tensile strain (right panel). Compressive strain closes the Dirac cone, further isolates the bands near the Fermi level, and maintains a nearly homogeneous density of states between the Dirac cone and the states slightly above the Fermi energy. In contrast, tensile strain opens the Dirac cone and enhances the DOS of the conduction bands near the Fermi level.} 
\end{figure*}

\begin{figure*}[h] 
    \centering
    \begin{minipage}[b]{0.48\textwidth}
        \centering
        \includegraphics[width=\textwidth]{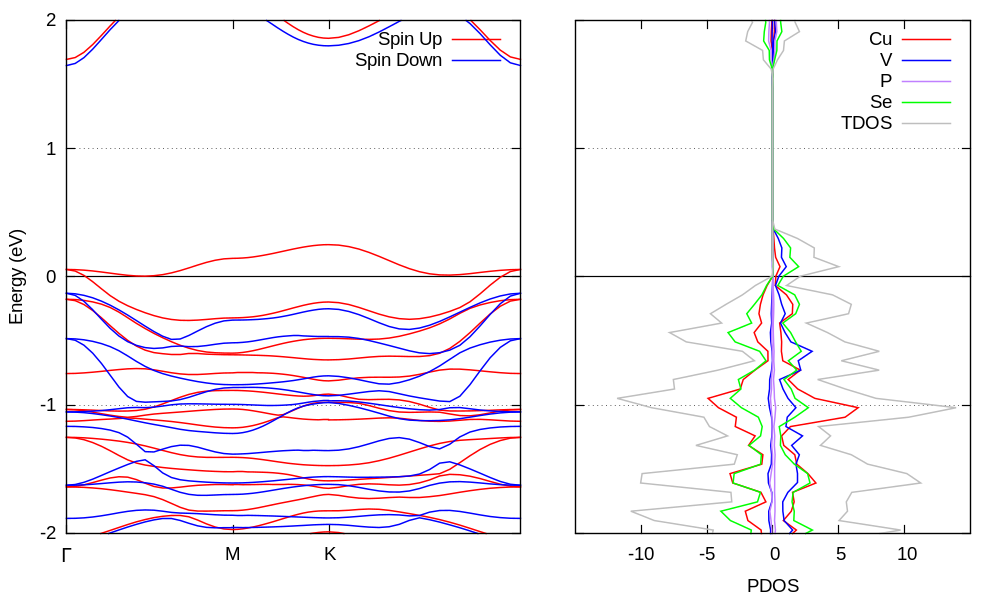}  
    \end{minipage}
    \hfill 
    \begin{minipage}[b]{0.48\textwidth}
        \centering
        \includegraphics[width=\textwidth]{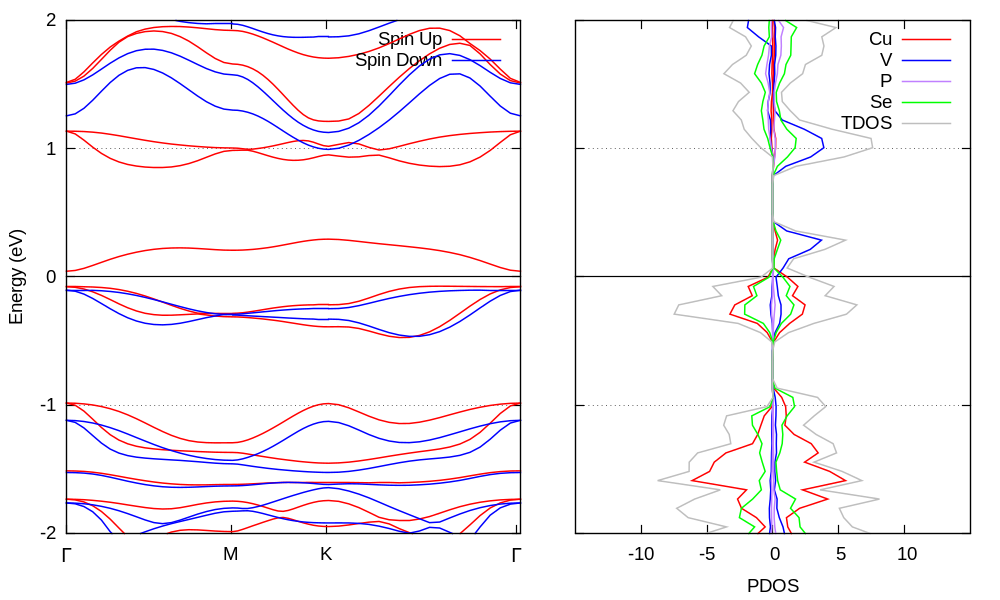} 
    \end{minipage}
    \caption{Band structure and projected density of states (PDOS) of the ferroelectric (FE) phase for $U=4$ (left panel) and the paraelectric (PE) phase for $U=4$ (right panel). In the FE phase, the inclusion of the $U$ parameter increases the overlap between valence bands, making the material more similar to a conventional metal rather than a half-metal in the absence of $U$. In contrast, in the PE phase, the $U$ parameter opens a small gap at the Fermi level, rendering the system semiconducting.} 
    \label{fig:PDOS_U-value}
\end{figure*}

\begin{figure*}[h] 
    \centering
    \begin{minipage}[b]{0.48\textwidth}
        \centering
        \includegraphics[width=\textwidth]{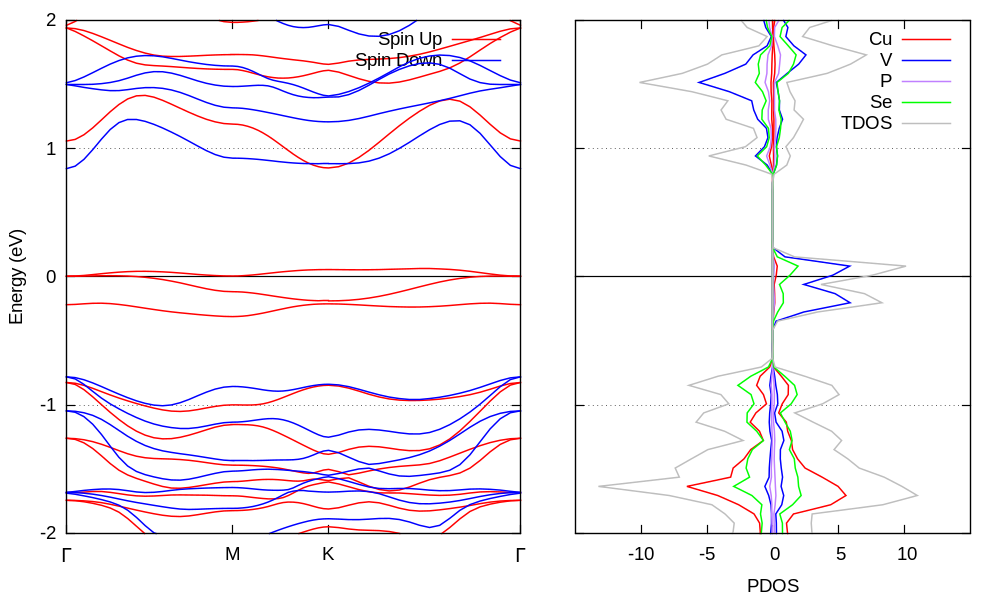} 
    \end{minipage}
    \hfill 
    \begin{minipage}[b]{0.48\textwidth}
        \centering
        \includegraphics[width=\textwidth]{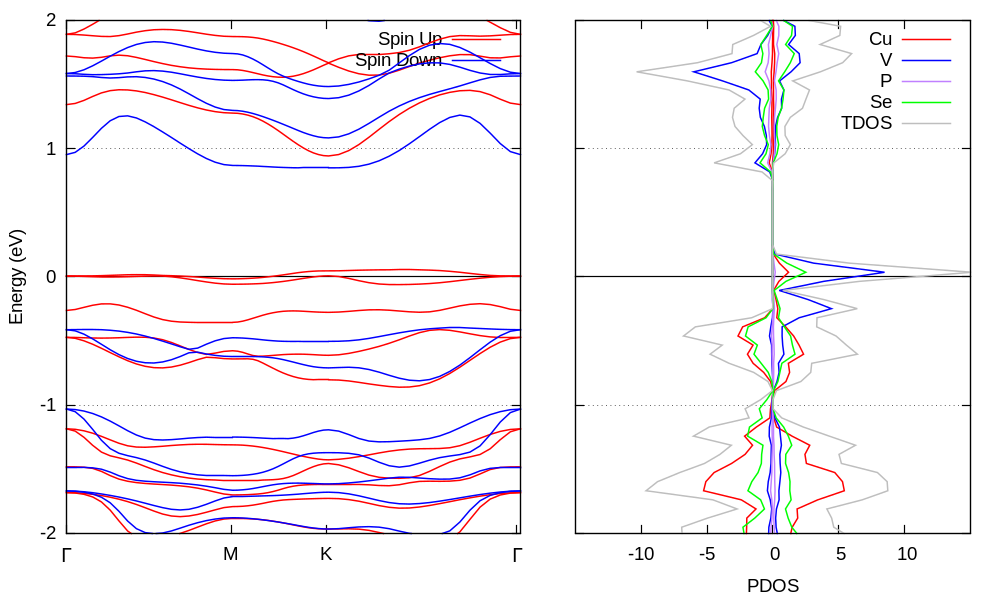}
    \end{minipage}
    \caption{Band structure and projected density of states (PDOS) of the ferroelectric (FE) stable phase at the energy minimum (left panel) and the FE unstable phase at the energy maximum (right panel). The sharpest peak in the density of states at the Fermi level occurs at the unstable FE configuration, corresponding to the Cu located half-way in the FE–PE phase transition.} 
    \label{fig:PE_FE_PDOS}
\end{figure*}

\begin{figure*}[h]
\section{Topological Properties}\label{tp}
\centering
\includegraphics[width=\textwidth]{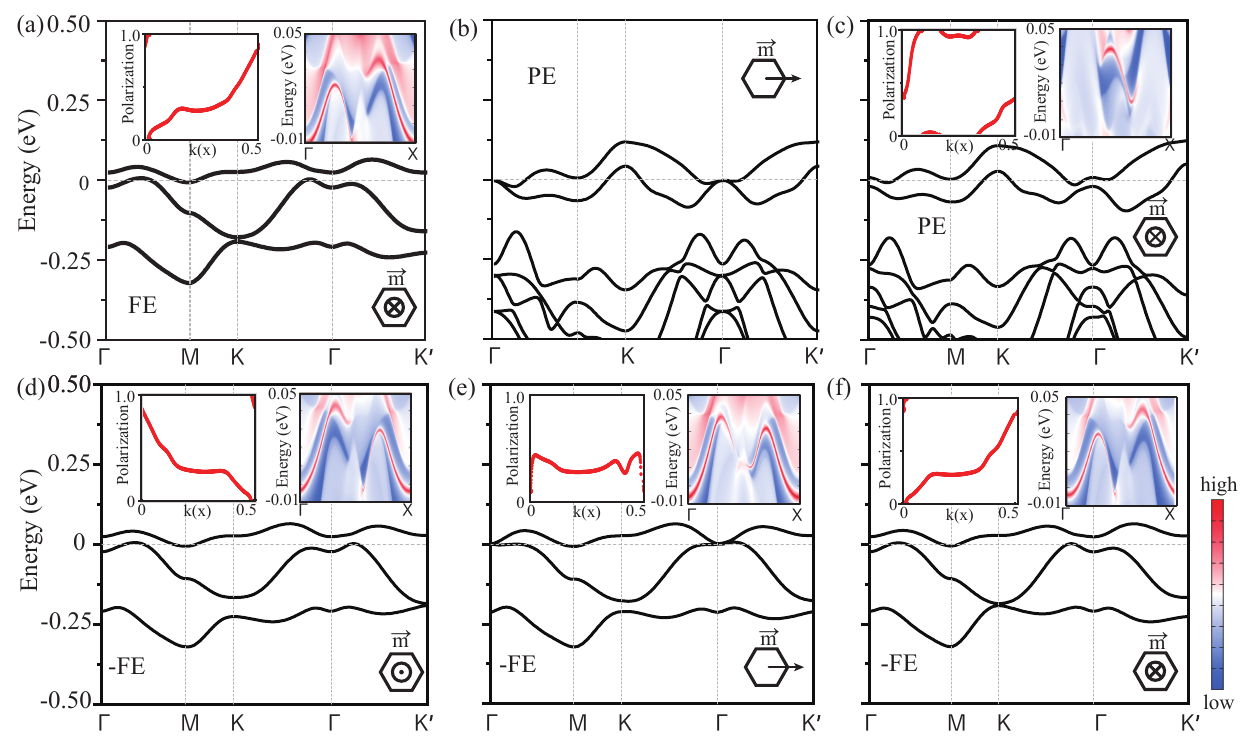} 
\caption{The supplement to the topological properties of the CuVP$_2$Se$_6$.
(a) FE phase with $-m_z$;
(b) PE phase with $m_x$;
(c) PE phase with $-m_z$;
(d) -FE phase with $m_z$;
(e) -FE phase with $m_x$;
(f) -FE phase with $-m_z$.
 The insets  represent the sum of the two occupied bands wannier charge centers (wcc) and edge states  of the corresponding magnetic configurations.}
\end{figure*}

\bibliography{Ref}

\end{document}